\documentclass[
 reprint,superscriptaddress,
 amsmath,amssymb,
 aps,
prb,
]{revtex4-2}
\usepackage{graphicx}
\usepackage{dcolumn}
\usepackage{bm}
\usepackage[normalem]{ulem} 
\DeclareTextCommand{\DJ}{OT1}{%
  \raisebox{-0.1ex}{\scalebox{0.75}[1.4]{--}}\kern-.4em D%
}
\usepackage{xcolor}
\usepackage{subfigure}

\begin{document}


\title{Reexamining the strange metal charge response with transmission inelastic electron scattering}

\author{Niels de Vries}
\email{niels@illinois.edu}
\affiliation{
Materials Research Laboratory, Grainger College of Engineering, University of Illinois, Urbana-Champaign, 104 S. Goodwin Ave., Urbana, IL, 61801, USA
}
\affiliation{
Department of Physics, Grainger College of Engineering, University of Illinois, Urbana-Champaign, 1110 W. Green St., Urbana, IL, 61801, USA
}
\author{Eric Hoglund}
\email{hoglunder@ornl.gov}
\affiliation{
Center for Nanophase Materials Sciences, Oak Ridge National Laboratory, Oak Ridge, TN 37830, USA
}
\author{Dipanjan Chaudhuri}
\affiliation{
Materials Research Laboratory, Grainger College of Engineering, University of Illinois, Urbana-Champaign, 104 S. Goodwin Ave., Urbana, IL, 61801, USA
}
\affiliation{
Department of Physics, Grainger College of Engineering, University of Illinois, Urbana-Champaign, 1110 W. Green St., Urbana, IL, 61801, USA
}
\author{Sang hyun Bae}
\affiliation{
Materials Research Laboratory, Grainger College of Engineering, University of Illinois, Urbana-Champaign, 104 S. Goodwin Ave., Urbana, IL, 61801, USA
}
\affiliation{
Department of Materials Science and Engineering, Grainger College of Engineering, University of Illinois, Urbana-Champaign, 1110 W. Green St., Urbana, IL, 61801, USA
}
\author{Jin Chen}
\affiliation{
Materials Research Laboratory, Grainger College of Engineering, University of Illinois, Urbana-Champaign, 104 S. Goodwin Ave., Urbana, IL, 61801, USA
}
\affiliation{
Department of Physics, Grainger College of Engineering, University of Illinois, Urbana-Champaign, 1110 W. Green St., Urbana, IL, 61801, USA
}
\author{Xuefei Guo}
\affiliation{
Materials Research Laboratory, Grainger College of Engineering, University of Illinois, Urbana-Champaign, 104 S. Goodwin Ave., Urbana, IL, 61801, USA
}
\affiliation{
Department of Physics, Grainger College of Engineering, University of Illinois, Urbana-Champaign, 1110 W. Green St., Urbana, IL, 61801, USA
}
\author{David Ba\l ut}
\affiliation{
Materials Research Laboratory, Grainger College of Engineering, University of Illinois, Urbana-Champaign, 104 S. Goodwin Ave., Urbana, IL, 61801, USA
}
\affiliation{
Department of Physics, Grainger College of Engineering, University of Illinois, Urbana-Champaign, 1110 W. Green St., Urbana, IL, 61801, USA
}
\author{Genda Gu}
\affiliation{
Division of Condensed Matter Physics and Materials Science, Brookhaven National Laboratory, Upton, NY, USA
}
\author{Pinshane Huang}
\affiliation{
Materials Research Laboratory, Grainger College of Engineering, University of Illinois, Urbana-Champaign, 104 S. Goodwin Ave., Urbana, IL, 61801, USA
}
\affiliation{
Department of Materials Science and Engineering, Grainger College of Engineering, University of Illinois, Urbana-Champaign, 1110 W. Green St., Urbana, IL, 61801, USA
}
\author{Jordan Hachtel}
\affiliation{
Center for Nanophase Materials Sciences, Oak Ridge National Laboratory, Oak Ridge, TN 37830, USA
}
\author{Peter Abbamonte}
\email{abbamont@illinois.edu}
\affiliation{
Materials Research Laboratory, Grainger College of Engineering, University of Illinois, Urbana-Champaign, 104 S. Goodwin Ave., Urbana, IL, 61801, USA
}
\affiliation{
Department of Physics, Grainger College of Engineering, University of Illinois, Urbana-Champaign, 1110 W. Green St., Urbana, IL, 61801, USA
}

\date{\today}

\begin{abstract}
The strange metal remains one of the great unsolved problems for 21st century science. Since the early development of the marginal Fermi liquid phenomenology, it has been clear that progress requires detailed knowledge of the momentum- and frequency-dependent charge susceptibility, $\chi(\mathbf{q},\omega)$, particularly at large momenta. Electron energy-loss spectroscopy (EELS), performed in either reflection or transmission geometry, provides the most direct probe of $\chi(\mathbf{q},\omega)$. However, measurements over the past four decades have yielded conflicting results, with some studies reporting a dispersing RPA-like plasmon and others observing a strongly overdamped, incoherent response. Here we report a transmission EELS study of Bi$_2$Sr$_2$CaCu$_2$O$_{8+x}$ (Bi-2212) that simultaneously achieves high energy resolution ($\Delta E \approx 30$~meV) and high momentum resolution ($\Delta q \approx 0.01$~\AA$^{-1}$). 
To address issues of reproducibility, measurements were repeated ten times on five different Bi-2212 flakes, benchmarked against aluminum, a well-characterized Fermi liquid, and quantitatively compared with prior studies spanning four decades.
At momenta $q < 0.15$~\AA$^{-1}$, we observe a highly damped plasmon whose linewidth is comparable to its energy. At larger momenta, $q > 0.15$~\AA$^{-1}$, this excitation does not disperse but instead evolves into an incoherent continuum, with no evidence for the RPA-like dispersion reported in some earlier works. Comparison with recent RIXS measurements on Bi-based cuprates supports the view that Bi-2212 is an incoherent metal with strongly damped charge excitations.
\end{abstract}

\maketitle

\section{\label{sec:introduction}Introduction}

The strange metal is a state of matter thought to represent a fundamental limit on the degree of quantum entanglement possible in a many-body system \cite{Zaanen2019,Zaanen2021,Hartnoll2022,Phillips2022}. First identified in the high-temperature metallic phase of the cuprate superconductors, the strange metal has since been observed in a broad range of systems, including materials that are not superconducting at all. Notable examples include the Bechgaard salts \cite{Doiron2009}, $\beta$-YbAlB$_4$ \cite{Tomita2015}, Ge-doped YbRh$_2$Si$_2$ \cite{Trovarelli1999}, bilayer Sr$_3$Ru$_2$O$_7$ \cite{Mousatov2020}, CeCoIn$_5$ \cite{Tanatar2007}, Ba(Fe$_{1-x}$Co$_x$)$_2$As$_2$ \cite{Doiron2009}, La$_3$Ni$_2$O$_{7-\delta}$ \cite{Zhou2025}, magic-angle bilayer graphene \cite{Cao2020}, and even certain ultracold Fermi gases in optical lattices \cite{Brown2018}.

For nearly forty years, the strange metal has been one of the central obstacles to understanding high-temperature superconductivity (HTSC) \cite{Keimer2015}. Decades of research have shown that, apart from their $d$-wave order parameter, the superconducting properties of the cuprates are not particularly exotic: they form singlet Cooper pairs, obey Ginzburg-Landau phenomenology for type-II superconductors, and exhibit standard features such as Andreev reflection and a zero-frequency pole in the complex conductivity $\sigma(\omega)$ \cite{Basov2005}. What is lacking is an agreed microscopic description of the high-temperature metallic phase from which superconductivity emerges.

The primary challenge for HTSC, and a broad class of other strongly correlated materials, is therefore to understand the strange metal itself. The inability to account for this state, whose relevance extends far beyond superconductivity, has been described as an intellectual crisis \cite{Zaanen2021}.

The phenomenology of strange metals differs fundamentally from that of ordinary metals. The Drude scattering rate in normal metals, which arises from quasiparticle collisions, is constrained by phase-space considerations to have the form \cite{Ashcroft1976,Stricker2014,Behnia2015}
\begin{equation}
\label{eq:drudescatteringrate}
\hbar\tau^{-1} \propto (\hbar\omega)^2 + (k_B T)^2 ,
\end{equation}
where $T$ is the temperature and $\omega$ is the frequency of the  current. The proportionality constant is highly material dependent, typically being smaller in high-carrier-density metals. Behavior consistent with Eq. 1 is observed in a wide range of garden variety metals including Cu, Ag, and Au \cite{Lawrence1976}, graphite, Bi \cite{Behnia2022}, and several oxides such as $n$-type SrTiO$_3$ \cite{vdMarel2011}, SrRuO$_3$ and Sr$_2$RuO$_4$ \cite{MackenzieMaeno2003,Stricker2014}. At sufficiently high temperatures, the scattering rate of normal metals saturates at the Mott-Ioffe-Regel (MIR) limit, where the mean free path approaches a microscopic length scale such as the lattice parameter or inverse Fermi momentum \cite{Hussey2004}.

By contrast, in strange metals the usual phase-space constraints appear not to apply. Instead, the scattering rate takes the form
\begin{equation}
\label{eq:strangescatteringrate}
\hbar\tau^{-1} \propto \sqrt{(\hbar\omega)^2 + (k_B T)^2}.
\end{equation}
A key consequence is that the DC resistivity ($\omega = 0$), with $\rho \propto \tau^{-1}$, becomes linear in temperature, i.e., $\rho \propto k_B T / \hbar$. Remarkably, the proportionality constant in Eq.~2 is nearly universal, having a value of order unity across a wide variety of strange metals with little material dependence \cite{Bruin2013}. Moreover, the resistivity shows no feature at the Debye temperature, where one might expect 
phonon scattering to cause a change in slope. This behavior suggests that the scattering rate lies close to a fundamental limit, sometimes referred to as the Planckian bound \cite{Zaanen2019,Hartnoll2022}.

Strange metals also violate the MIR limit \cite{Hussey2004}, indicating a breakdown of the quasiparticle picture altogether. The prevailing view is that these properties are a result of strong interactions and a high degree of quantum entanglement in the underlying many-body state \cite{Hartnoll2022,Phillips2022,BalutStrange2025}.

There is currently no agreed microscopic theory of the strange metal. Explanations based on perturbative, quasiparticle methods have made little progress \cite{Zaanen2019,Phillips2022}. 
Many of the basic properties of strange metals are described by the marginal Fermi liquid (MFL) phenomenology, which is not a microscopic theory but a hypothesis about the functional form of the dynamic charge susceptibility \cite{Varma1989}. MFL has emphasized the importance of accurate measurement of the collective charge response in strange metals, particularly at nonzero momentum, $q$. 

Experimental efforts to measure the dynamic charge susceptibility, $\chi''(q,\omega)$, of strange metals have produced conflicting results \cite{Abbamonte2025}. The most direct probes of the charge response are infrared (IR) optics, which measures the response at zero momentum ($q=0$), and inelastic electron scattering in the form of electron energy-loss spectroscopy (EELS), which accesses finite momentum transfers ($q\neq0$) in either reflection or transmission geometries \cite{Abbamonte2025}. Resonant inelastic x-ray scattering (RIXS) probes a related quantity that mixes charge and spin responses \cite{Ament2011,vdBrink2007,Lomeli2025}.
The absence of a well-defined relationship between the RIXS cross section and a corresponding Green’s function limits its quantitative use for testing MFL and related ideas. 

Most experimental work has focused on the prototypical strange metal Bi$_2$Sr$_2$CaCu$_2$O$_{8+x}$ (Bi-2212), as it cleaves easily. We begin with a chronological overview of prior IR and EELS studies on this material, summarized as impartially as possible.

\begin{figure*}
\includegraphics[width=0.8\textwidth]{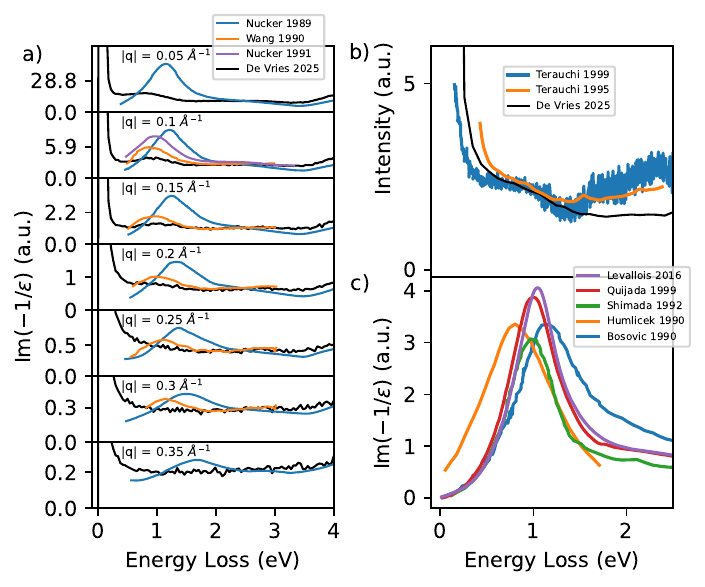}
\caption{\label{fig:measurements_overview}(a) Loss function of Bi-2212 at various momenta, measured with momentum-resolved T-EELS as reported in Ref.~\cite{Nucker1989} (blue line), Ref.~\cite{Wang1990} (orange line), Ref.~\cite{Nucker1991} (purple line), compared to our current study (black line). (b) Low-energy loss function of Bi-2212 measured with momentum-integrated T-EELS from Refs.~\cite{Terauchi1999} (blue line) and ~\cite{Terauchi1995} (orange line) compared to our current study (black line). (c) Loss function of Bi-2212 at $q=0$ extracted from several representative IR optics studies \cite{Bozovic1990,Humlicek1990,Shimada1992,Quijada1999,Levallois2016}.}
\end{figure*}

\section{Previous studies of the charge response of Bi-2212}

The first reported measurement of the dynamic charge response in Bi-2212 was carried out using a dedicated monochromated transmission EELS (T-EELS) spectrometer at the Karlsruhe Institute of Technology, with an energy resolution of 150 meV \cite{Nucker1989}. Although the instrument achieved excellent momentum resolution (0.04 \AA$^{-1}$), its spatial resolution was limited to several millimeters \cite{Fink1985}. The published spectra revealed a series of interband transitions spanning energies from 5-35 eV, as well as a prominent plasmon peak near 1 eV. The plasmon dispersion was consistent with Lindhard theory within the random phase approximation (RPA) \cite{Pines2018} and was observed up to the largest measured momentum, $q = 0.35$ \AA$^{-1}$. These data are reproduced in Fig.~\ref{fig:measurements_overview}(a), labeled “Nücker 1989.”

Focusing on the low-energy portion of the spectrum (0–2 eV), infrared optics studies of Bi-2212 at $q \approx 0$ likewise reported a plasmon excitation that closely resembles the feature observed in T-EELS at the smallest measured momentum \cite{Bozovic1990,Humlicek1990}, albeit at a somewhat different energy. The loss functions extracted from these infrared measurements are shown in Fig.~\ref{fig:measurements_overview}(c), labeled “Bozovic 1990” and “Humlicek 1990.”

\begin{figure}
\includegraphics[width=0.98\columnwidth]{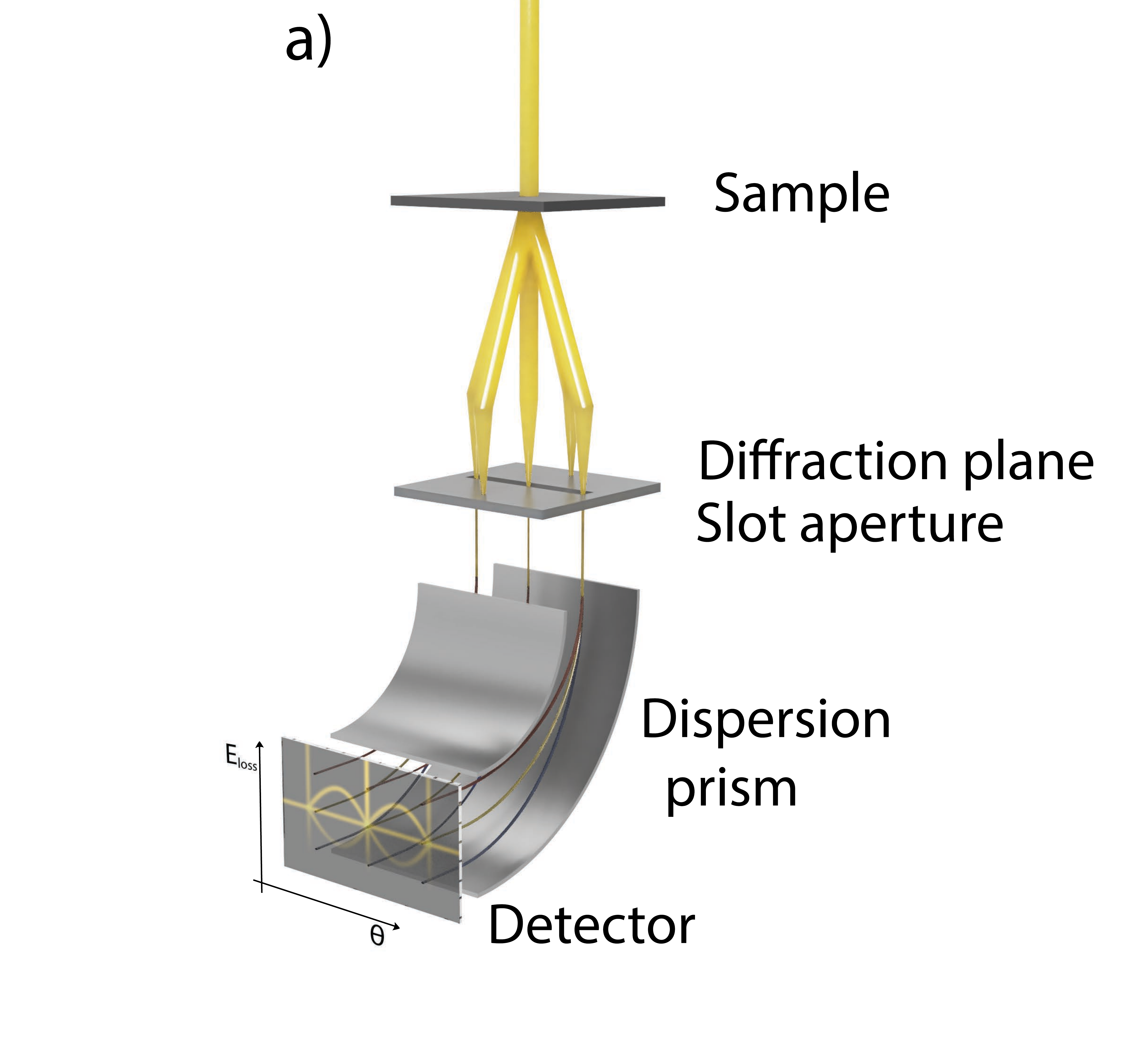}
\includegraphics[width=0.98\columnwidth]{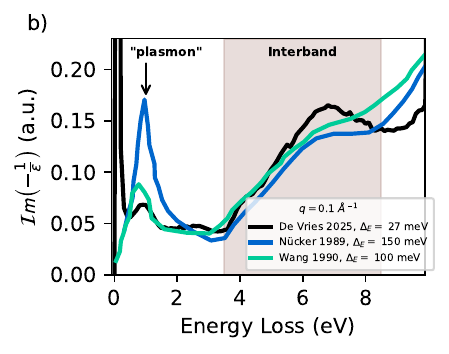}
\caption{\label{fig:schematic}(a) Schematic representation of a T-EELS experiment. The momentum selection is made with a slot aperture in the diffraction plane, after which the beam is dispersed in the spectrometer. (b) The loss function $\mathrm{Im}(-1/\epsilon(\omega))$ at $q=0.1$~\AA$^{-1}$, measured with EELS as previously published in Refs. \cite{Nucker1989,Wang1990}, and as measured in this study. The legend lists the energy resolution in each experiment, as given by the FWHM of the quasielastic line. }
\end{figure}

The next T-EELS study of Bi-2212 was carried out at Virginia Tech with slightly improved energy resolution (100~meV) and comparable momentum resolution (0.04~\AA$^{-1}$) \cite{Wang1990}. Although the authors claimed consistency with the Karlsruhe results, the measured dispersion and lineshape of the $\sim$1~eV excitation were quite different. These data are also shown in Fig.~\ref{fig:measurements_overview}(a) for comparison, labeled “Wang 1990.” At higher energies ($E_{\mathrm{loss}} > 3.5$~eV), the spectra were dominated by interband transitions, which qualitatively agree with the Karlsruhe measurements, as illustrated in Fig.~\ref{fig:schematic}(b).

The following year, the Karlsruhe group repeated their T-EELS experiment at a fixed momentum of $q = 0.1$~\AA$^{-1}$, using similar energy and momentum resolutions as in their original study (130~meV and 0.05~\AA$^{-1}$, respectively) \cite{Nucker1991}. As shown in Fig.~\ref{fig:measurements_overview}(a) and labeled “Nücker 1991,” this one spectrum agrees more closely with the Virginia Tech measurements \cite{Wang1990} than with the group’s own earlier results \cite{Nucker1989}.

In 1992, an additional infrared optics study at $q \sim 0$ reported a peak near 1~eV, albeit at a somewhat different energy than earlier optical measurements on Bi-2212 \cite{Shimada1992}. This result is included in Fig.~\ref{fig:measurements_overview}(c), labeled “Shimada 1992.”

More recent infrared studies published in 1999 \cite{Quijada1999} and 2016 \cite{Levallois2016} again observed a well-defined plasmon in Bi-2212, with a plasma frequency of approximately 1~eV. The corresponding loss functions are shown in Fig.~\ref{fig:measurements_overview}(c), labeled “Quijada 1999” and “Levallois 2016,” respectively.

It is important to note that Ref.~\cite{Levallois2016} reported an unexpectedly weak doping dependence of the 1~eV plasmon in Bi-2212 across samples with superconducting transition temperatures ranging from $T_c = 60$~K on the underdoped side to $T_c = 58$~K on the overdoped side. Over this range, the hole concentration in Bi-2212 varies from $p = 0.11$ to $p = 0.24$---more than a factor of two \cite{Drozdov2018}. In a conventional metal, such a change in carrier density would be expected to result in a plasma frequency shift of more than 40\% \cite{Ashcroft1976}. Instead, the observed shift was less than 5\%, making it difficult to understand the 1~eV excitation as a plasmon in the usual sense. Moreover, this result rules out variations in carrier density as the origin of the differing plasma frequencies shown in Fig.~\ref{fig:measurements_overview}(c), suggesting that those discrepancies instead arise from other forms of experimental uncertainty.

The early T-EELS studies were challenged in the second half of the 1990's by a group at Tohoku University, who developed a monochromated transmission electron microscope (TEM) equipped with an EELS detector. This instrument achieved vastly improved energy resolution, albeit with poorer momentum resolution than previous T-EELS instruments \cite{Nucker1989,Wang1990,Nucker1991}. In their first study of Bi-2212, performed with an energy resolution of 39~meV and an effective momentum resolution of $\Delta q \approx 0.3$~\AA$^{-1}$, no plasmon feature near 1~eV was observed at all \cite{Terauchi1995}. Instead, over the accessible momentum range $q \lesssim 0.3$~\AA$^{-1}$, the low-energy response consisted of a featureless continuum forming a shoulder on the elastic line.

A subsequent experiment with improved energy resolution, $\Delta E=$ 25~meV and $\Delta q \approx 0.3$~\AA$^{-1}$, again failed to detect a 1~eV plasmon in Bi-2212 \cite{Terauchi1999}. Both Tohoku measurements are shown in Fig.~\ref{fig:measurements_overview}(b), labeled “Terauchi 1995” and “Terauchi 1999,” respectively.

Most recently, the charge response of Bi-2212 has been investigated using reflection momentum-resolved EELS (M-EELS) \cite{Mitrano2018}. These measurements achieve an energy resolution of 4~meV and a momentum resolution of 0.02~\AA$^{-1}$ \cite{Mitrano2018,Husain2019,Chen2024}. The results are quantitatively consistent with earlier reflection HR-EELS measurements, which stated an energy resolution of 30~meV and momentum resolution of 0.01~\AA$^{-1}$ \cite{Schulte2002}. Moreover, in the $q \to 0$ limit, the M-EELS data were shown to agree quantitatively with infrared optics measurements \cite{Chen2024,Levallois2016}.

Despite this consistency, these reflection EELS results appear to conflict with the early transmission EELS studies, as a plasmon is observed only at small momenta ($q \lesssim 0.1$~\AA$^{-1}$) and not at larger values of $q$ \cite{Mitrano2018,Husain2019,Chen2024}.

In summary, reflection EELS, transmission EELS, and infrared optics experiments broadly agree that Bi-2212 exhibits a plasmon-like excitation near 1~eV in the long-wavelength limit, albeit with some variation in energy and lineshape. However, significant discrepancies remain in the reported behavior of the charge response at finite momentum, with some measurements seeing a well-defined plasmon and others observing no peak at all. 

These discrepancies in the measured charge response of Bi-2212 are deeply concerning. The strange metal remains one of the great unsolved problems for 21st century science. Understanding its origin, and whether its phenomenology adheres to a marginal Fermi liquid picture or something else, ultimately depends on an accurate knowledge of the dynamic charge susceptibility, $\chi''(q,\omega)$, at finite momentum, $q$. The inconsistency of experimental results therefore poses a serious problem. Some superficial difference between reflection M-EELS and transmission EELS data is expected because M-EELS probes a surface response whose connection to the bulk susceptibility has only recently been quantified \cite{Chen2024}. However, mutually inconsistent T-EELS measurements on the same material represent the core impediment to progress on this critical problem.  

At the moment, there are at least three possible factors that might explain the varying outcomes of different measurements. 

The first is the effect of finite energy resolution. The first T-EELS measurements on Bi-2212 \cite{Nucker1989,Wang1990,Nucker1991} were performed with energy resolutions of 130~meV or larger. In these early experiments, monochromation was performed prior to acceleration, which leads to a convolution of the intrinsic spectral lineshape of the monochromated electron beam with the noise distribution of the high voltage power supply. This results in a Voigt-like energy resolution function exhibiting long spectral tails (modern EELS instruments partially mitigate this issue by monochromating the electron beam at ground potential, producing a Gaussian spectral profile \cite{Krivanek2013, Dellby2023}). One consequence is that the zero-loss peak obscures low-energy spectral features, a complication explicitly noted in several of the early studies \cite{Nucker1989,Wang1990}. The degree to which the zero-loss peak obscures low-energy excitations would be substantially reduced in experiments performed with much higher energy resolution, such as the 25~meV measurements of Ref.~\cite{Terauchi1999} or the 5~meV measurements enabled by modern instruments \cite{Mitrano2018, Husain2019, Chen2024}.

A second possible factor is differences in data processing. In the earliest T-EELS studies, the published spectra were obtained by subtracting a fit to the zero-loss peak and applying multiple-scattering corrections to the raw data \cite{Nucker1989, Wang1990, Nucker1991}. Neither the raw spectra nor a description of how these corrections affected the results were disclosed, making it impossible to assess their effect. Later, high-energy-resolution measurements by the Tohoku group, whose raw data do not show a plasmon feature, demonstrated that applying similar subtraction and correction procedures can artificially generate a peak near 0.6~eV \cite{Terauchi1995}.  

A third possibility is differences in momentum resolution. While the studies in Refs.~\cite{Terauchi1995, Terauchi1999} had substantially better energy resolution than the early T-EELS experiments \cite{Nucker1989, Wang1990, Nucker1991}, their momentum resolution of $\Delta q \sim 0.3$~\AA$^{-1}$ is significantly worse. As a result, these measurements effectively integrated over a significant portion of the plasmon dispersion curve, which could broaden and possibly obscure the excitation. 
Nevertheless, the plasmon peak in Ref. \cite{Nucker1989} is clearly visible all the way up to $q = 0.35$~\AA$^{-1}$. Therefore, even with this degree of momentum averaging, a well-defined plasmon peak should still have been observable in the data \cite{Husain2021Reply}. 

\section{Aim of this work}
In light of the above inconsistencies, it is necessary to revisit transmission EELS measurements of the finite-$q$ charge response of strange metals, beginning with Bi-2212, with a focus on achieving simultaneous high energy and momentum resolution while minimizing data processing and corrective procedures. Over the past several decades, EELS instrumentation has undergone transformational advances, making it possible to attain meV-scale energy resolution together with high momentum resolution in collimated-beam STEM mode \cite{Krivanek2019, Song2025, Colliex2022}. In parallel, sample preparation techniques for transmission electron microscopy have improved dramatically to meet the needs of aberration-corrected, high-resolution scanning TEM (STEM) studies with sub-\AA~spatial resolution. Developments driven by two-dimensional materials research have further enabled unprecedented control over crystalline flakes down to monolayer thickness, along with highly reproducible preparation protocols. At the same time, advances in detector efficiency now allow measurements at substantially lower electron doses, reducing beam-induced damage while improving sensitivity to weak signals. Finally, the quality and homogeneity of cuprate single crystals have improved significantly over the past two decades \cite{Erb2015}. 

The aim of this work is to take advantage of these advances to obtain new measurements of the dynamic charge response of strange metals using transmission EELS (T-EELS) with state-of-the-art instrumentation and sample preparation, with an emphasis on reproducibility and comparison to control materials whose charge response is known and can be used to check the operating conditions of the instrument. By revisiting this problem with substantially improved energy and momentum resolution, and minimal data processing, we seek to provide a fresh experimental perspective on a longstanding controversy that is critical for future understanding of Planckian dissipation and strange metals.


\section{Materials and methods}
In this study, we performed transmission EELS (T-EELS) measurements on optimally doped, strange metal phase Bi-2212. Single crystals were grown at Brookhaven National Laboratory using methods described previously \cite{WenGrowth2008}. The superconducting transition temperature, $T_c = 92$~K, was confirmed using a Quantum Design MPMS SQUID magnetometer. Prior to T-EELS measurements, the crystallographic orientation of each crystal was determined by Laue diffractometry. The crystals were then exfoliated using Scotch tape and transferred onto holey SiN TEM grids using PDMS and silicon, with the orientation of the crystals tracked throughout the process. The size and thickness of the resulting flakes were checked using optical dark-field microscopy. The grids were stored under high vacuum overnight before being loaded into the electron microscope. Ten distinct T-EELS measurements were done on five different Bi-2212 flakes, summarized in Table~\ref{table:samples}. 

To validate our measurements, and to check for potential experimental artifacts, we performed identical T-EELS measurements on a control sample of high-purity, polycrystalline aluminum under the same operating conditions. Aluminum is a well-established Fermi-liquid metal whose charge response has been extensively characterized by both EELS and x-ray techniques \cite{Tischler2003,Batson1982, Batson1983,Platzman1992,Schulke1993}, making it a suitable reference material. The aluminum sample was a commercially obtained diffraction standard, consisting of a 30~nm layer of aluminum evaporated onto carbon film. Prior to insertion into the microscope, the specimen was baked at 100 °C under high vacuum to remove surface contamination.

The T-EELS measurements on both Bi-2212 and aluminum were carried out at the Center for Nanophase Materials Sciences at Oak Ridge National Laboratory using a Nion HERMES monochromated, aberration-corrected scanning transmission electron microscope (STEM) equipped with a Dectris ELA hybrid pixelated area detector. All experiments were performed at room temperature with a beam energy of 60~keV in transmission geometry with the beam incident along the (0,0,1) zone axis. In this geometry, the momentum transfer lies parallel to the CuO$_2$ planes, i.e., $q_z \sim 0$. Condenser and monochromator aberrations were corrected using a 30~mrad reference configuration before switching the instrument to a three-condenser TEM mode. The beam convergence was reduced to 50~$\mu$rad or less, corresponding to a momentum spread of $\lesssim 0.01$~\AA$^{-1}$ in vacuum, and yielding a probe diameter of approximately 200~nm. The exact momentum resolution varied between runs, as summarized in  Table~\ref{table:samples}. 

T-EELS intensity maps, $I(q,\omega)$, were acquired by projecting the diffraction pattern into the spectrometer through a 125$\times$2200~$\mu$m slot aperture, as  illustrated in Fig.~\ref{fig:schematic}(a).
Spectrometer aberrations were corrected up to
third order, yielding an energy resolution of 30 meV or better in vacuum, as determined from the full width at half maximum (FWHM) of the bright-field peak in vacuum.
Data were collected as a series of 1200-2400 frames, each with an exposure time of 500~ms. The momentum-space width of the slot aperture was calibrated by acquiring non-dispersed diffraction images with and without the slot inserted.

Ten distinct T-EELS measurements were performed on
five different Bi-2212 crystals, summarized in Table~\ref{table:samples}, to assess reproducibility. In these measurements, the mean energy resolution after transmission through the sample was 38~meV, as determined from the full width at half maximum (FWHM) of the zero-loss peak, and the mean momentum resolution in the non-dispersive direction was 0.014~\AA$^{-1}$ (see Table~\ref{table:samples}). Measurements were carried out at multiple momenta along the $(h,h,0)$ or $(\bar{h},h,0)$ directions of the tetragonal unit cell, i.e., parallel and perpendicular to the supermodulation, respectively, as summarized in Table~\ref{table:samples}. In the table, the relative thicknesses of the various flakes are also given as the log-ratio of the thickness and the total inelastic mean free path\cite{Egerton2011}
\begin{equation}
\label{eq:toverlambda}
t/\lambda = \ln(I_t/I_0) .
\end{equation}
The quasielastic intensity $I_0$ was approximated by integrating the EELS signal using the trapezoid rule in the energy loss range between $-0.1$ and $0.1$~eV; the total intensity $I_t$ was obtained by integrating in the energy loss range from $-0.1$ to $1.8$~eV.

To obtain EELS spectra, offsets in individual detector frames were corrected by aligning to the bright field disk, which appears in the spectra as a quasielastic line. The momentum transfer was then computed for each detector pixel from the scattering angle and calibrated energy loss. Intensity curves at fixed momentum $q$ were obtained by binning the data within a fixed momentum range. For small momentum transfer $q<0.1$~\AA$^{-1}$, the bin width is taken to be 0.01~\AA$^{-1}$ and for momentum transfer $q\geq0.1$~\AA$^{-1}$, it is $0.02$~\AA$^{-1}$. In order to compute the loss function $Im(-1/\epsilon)$, the Coulomb matrix element, $|q|^{-2}$, and the Bose factor were divided out, i.e., 
\begin{equation}
\label{eq:lossfunction_crosssection}
\mathrm{Im}\left[-\frac{1}{\epsilon(q,\omega)}\right] \propto q^2 (1-e^{\hbar\omega/k_BT})\frac{d^2\sigma}{d\Omega d\omega} .
\end{equation}
No other corrections were performed on any of the data presented in this study.

We compare the results from this study to data presented in the previous works shown in Fig. 1 and Fig. 2(b). The data shown from other publications were extracted using WebPlotDigitizer~\cite{webplotdigitizer}.

\begin{table}
\begin{tabular}{r|ccccc}
Label & Flake & $\Delta_E$ (meV) & $\Delta_q$ (\AA$^{-1}$) & $t/\lambda$ & dir  \\
\hline
A & 1 & 32 & 0.008 & 0.09$\pm$0.01 & $\parallel$  \\
B & 1 & 32 & 0.014 & 0.09$\pm$0.01 & $\parallel$ \\
C & 1 & 37 & 0.007 & 0.09$\pm$0.01 & $\parallel$ \\
D & 2 & 27 & 0.024 & 0.02$\pm$0.01 & $\perp$ \\
E & 2 & 31 & 0.014 & 0.02$\pm$0.01 & $\perp$ \\
F & 3 & 64 & 0.012 & 0.02$\pm$0.01 & $\perp$ \\
G & 3 & 47 & 0.013 & 0.02$\pm$0.01 & $\perp$ \\
H & 4 & 37 & 0.021 & 0.02$\pm$0.01 & $\perp$ \\
I & 4 & 41 & 0.013 & 0.02$\pm$0.01 & $\perp$ \\
J & 5 & 31 & 0.014 & 0.02$\pm$0.01& $\perp$
\end{tabular}
\caption{\label{table:samples}Ten measurements performed on five different Bi-2212 flakes used for STEM-EELS measurements with associated energy resolution, momentum resolution, relative sample thickness, and the direction of $q$ defined with respect to the supermodulation direction.}
\end{table}

\section{Results}
\begin{figure}
\includegraphics[width=\columnwidth]{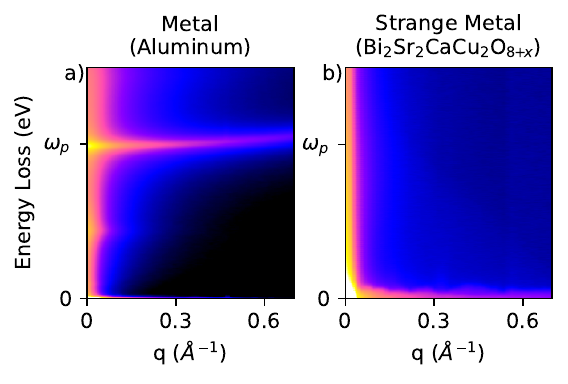}
\caption{\label{fig:metal_strangemetal_comparison} (a) T-EELS intensity $I(q,\omega)$ measured from aluminum, a classic Lindhard metal, as a function of $q$ and $\omega$, with the energy axis scaled to the plasma frequency, $\omega_P$=15 eV. Both the bulk and surface plasmon modes are visible, in agreement with many past studies. (b) T-EELS intensity $I(q,\omega)$ measured under the same conditions from the strange metal Bi-2212, this time scaled to $\omega_P$=1 eV. No similarly well-defined excitation is visible. This panel shows the sum of datasets F and G, as labeled in Table \ref{table:samples}.}
\end{figure}

We begin by comparing the raw T-EELS data for aluminum, a prototype Lindhard metal, and the strange-metal phase of Bi-2212. Fig.~\ref{fig:metal_strangemetal_comparison} shows color maps of the EELS intensity $I(q,\omega)$ as a function of $q$ and $\omega$, with the energy axes scaled by the nominal plasma frequency, $\hbar\omega_P \approx 15$ eV for aluminum and $\hbar\omega_P = 1$~eV for Bi-2212. The Bi-2212 shows the sum of datasets F and G as labeled in Table \ref{table:samples}.

The contrast between the two datasets is striking. Aluminum exhibits a sharp, well-defined plasmon with  textbook Lindhard-RPA dispersion, whose energy increases quadratically with $q$. A second, lower energy feature corresponds to a surface plasmon excited as the electron beam enters or exits the specimen. These features are consistent with numerous prior measurements of aluminum \cite{Batson1983} and confirm proper functioning of the instrument. By contrast, the $\sim$1 eV feature in Bi-2212 is far more subtle. Although a broad distribution of spectral weight appears near the plasma frequency at small $q$, no pronounced feature comparable to that in aluminum is discernible.

\begin{figure*}
\includegraphics[width=0.95\columnwidth]{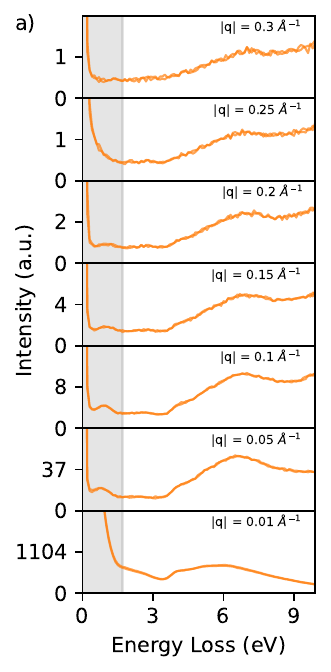}
\includegraphics[width=0.95\columnwidth]{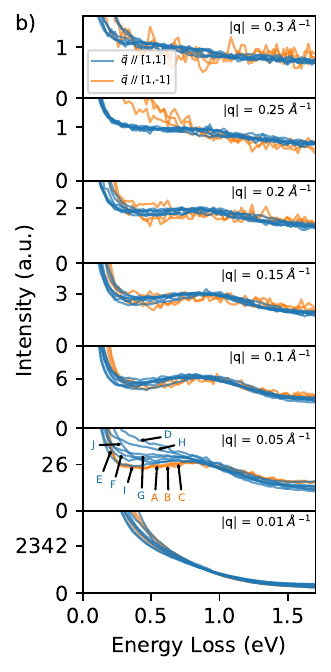}
\caption{\label{fig:bi2212-lineplots} 
Line plots of the T-EELS intensity of Bi-2212 measured at momenta between 0.01 and 0.25~\AA$^{-1}$. Blue and orange curves denote measurements taken perpendicular and parallel to the supermodulation direction, respectively. (a) Data from two experiments (A and B; see Table~\ref{table:samples}) measured over an energy range up to 10.5~eV. (b) T-EELS spectra in the 0-1.7~eV range (shaded region in panel (a)) for ten different measurements. Variations at small momentum transfer ($q = 0.05$~\AA$^{-1}$) are primarily due to differences in momentum resolution, while the large anisotropy at $q = 0.25$~\AA$^{-1}$ arises from a supermodulation Bragg peak at $q = 0.24$~\AA$^{-1}$.}
\end{figure*}

The key features in Bi-2212 are more clearly visualized in the line plots shown in Fig.~\ref{fig:bi2212-lineplots}, which display the T-EELS intensity as a function of $\omega$ at fixed momentum for the different measurements listed in Table~\ref{table:samples}. Each panel corresponds to a selected momentum in the range $q = 0.01$-$0.3$~\AA$^{-1}$, as indicated. 
These curves were obtained by binning the data from the detector (see Section IV), but otherwise are shown in their raw, unprocessed form, with the elastic line retained and without corrections for multiple scattering or other effects, to emphasize the intrinsic features in the measured spectra.

Figure~\ref{fig:bi2212-lineplots}(a) shows spectra from measurements A and B over the energy range 0-10~eV. Aside from the quasielastic line, the dominant features at all momenta are a broad series of peaks between 4 and 8~eV previously identified as interband transitions \cite{Terauchi1995}. At lower energies and small momenta ($q<0.1$~\AA$^{-1}$), subtle features are visible between 1.5 and 4~eV whose energy is comparable to the \textit{d}–\textit{d}$^{\prime}$ excitations reported in RIXS studies \cite{Barantani2022}, though their appearance only at small $q$ suggests a different origin.
Finally, a broad feature centered near 1~eV is observed over the momentum range $0.05$~\AA$^{-1} < q < 0.15$~\AA$^{-1}$, which we attribute to the plasmon-like mode. At smaller momenta ($q < 0.05$~\AA$^{-1}$), this feature appears only as a shoulder on the quasielastic line rather than as a distinct peak. With increasing momentum, its intensity decreases and the feature becomes unobservable for $q > 0.15$~\AA$^{-1}$.

Figure~\ref{fig:bi2212-lineplots}(b) focuses on the low-energy region (0-1.7 eV), corresponding to the shaded area in Fig.~\ref{fig:bi2212-lineplots}(a). Data from all ten measurements listed in Table~\ref{table:samples} are shown to assess the reproducibility. At the lowest momentum, $q=0.01$~\AA$^{-1}$, the spectra are dominated by the quasielastic peak due to proximity to the bright-field disk, which obscures inelastic features. At $q=0.05$~\AA$^{-1}$, a plasmon-like feature becomes visible, although the degree of interference from the elastic line varies strongly with the momentum resolution. In the measurements with the worst momentum resolution (i.e. D and H), there is no clear maximum. In the measurements with the best momentum resolution, a clear maximum is observed, consistent with a plasmon; however, its linewidth is comparable to its energy, indicating that it is only a marginally well-defined excitation. This feature persists at $q=0.1$~\AA$^{-1}$ and $q=0.15$~\AA$^{-1}$, but shows no clear dispersion, in contrast to expectations for a Fermi-liquid plasmon (see, e.g., Fig.~\ref{fig:metal_strangemetal_comparison}(a)). For momenta $q>0.15$~\AA$^{-1}$, the peak is no longer discernible and the spectra evolve into a broad continuum. This behavior differs markedly from Ref.~\cite{Nucker1989}, which reported a dispersing plasmon persisting up to momenta $q=0.45$~\AA$^{-1}$ or higher.

We now make a quantitative comparison between our measured loss function and that reported in previous previous works. Fig. \ref{fig:schematic}(b) shows the loss functions over the range 0-10~eV taken at $q=0.1$~\AA$^{-1}$ from the Karlsruhe group (1989) \cite{Nucker1989}, the Virginia group (1990) \cite{Wang1990}, and this work, labeled ``N\"ucker 1989," ``Wang 1990," and ``De Vries 2025," respectively. These measurements were performed under comparable  conditions and therefore provide a useful check of reproducibility.
For loss energies greater than 3.5~eV, the loss function is dominated by interband transitions, marked by the shaded region. In this range, all loss functions show reasonable agreement. 

In the region below 1.5 eV, which is relevant to the plasmon, there is substantial disagreement among the experiments. The spectrum labeled ``N\"ucker 1989" exhibits a strong, sharp peak at 1~eV, while ``Wang 1990" shows a broader and less pronounced peak centered near 0.8 eV. Our present measurements reveal an even weaker and broader feature around 1~eV. In our data, the spectrum below 0.3 eV is dominated by quasielastic scattering arising from our finite energy resolution, $\Delta E \sim 39$~meV. Although the ``N\"ucker 1989” and ``Wang 1990" spectra were acquired with even coarser resolutions (150 meV and 100 meV, respectively), no quasielastic feature is visible in those data, which were instead extrapolated to zero intensity in this energy range.


A more focused comparison of the plasmon region is shown in Fig.~\ref{fig:measurements_overview}(a), which compares our loss functions in the 0-4~eV range for several values of $q$ with those reported in earlier studies discussed in Section~I. All spectra are normalized by the integrated spectral weight between 2 and 3~eV, a range over which the loss functions exhibit minimal qualitative differences.
At small momentum, $q = 0.05$~\AA$^{-1}$, our loss function displays only a weak shoulder on the tail of the quasielastic line, whereas ``N\"ucker 1989" \cite{Nucker1989} shows a well-defined peak at a higher energy loss.
At energy losses below 0.5~eV, the spectra differ qualitatively, which we attribute to the removal of the quasielastic feature in earlier measurements \cite{Nucker1989}. 

A similar disagreement is observed at $q = 0.1$~\AA$^{-1}$. However, at this momentum, two additional datasets are available for comparison. The ``N\"ucker 1991" loss function shows improved agreement with our data, both in its energy and lineshape. Agreement with the data of ``Wang 1990" is better still, with near overlap of the corresponding features. As before, at energies below 0.5~eV the curves diverge significantly, reflecting the removal of the elastic line in earlier studies.

At larger momenta, $q > 0.15$~\AA$^{-1}$, the disagreement between datasets becomes pronounced. In our measurements, no plasmon feature is observed in this momentum range for any of the ten measurements (Fig.~\ref{fig:measurements_overview}(a)). By contrast, both ``N\"ucker 1989" and ``Wang 1990" report clear plasmon peaks at these momenta, although the reported features differ substantially from each other in energy, lineshape, and dispersion. As before, significant discrepancies at low energy arise from the removal of the elastic line in these earlier studies.

In Figs.~\ref{fig:measurements_overview}(b), we compare our results with the high-energy-resolution, momentum-integrated studies of ``Terauchi 1999" \cite{Terauchi1999} and ``Terauchi 1995" \cite{Terauchi1995}, which achieved energy resolutions of 25~meV and 39~meV, respectively. Although these studies achieved substantially higher energy resolution than earlier work, their momentum resolution was comparatively poor---0.3~\AA$^{-1}$ in both the 1995 and 1999 studies. To enable a direct comparison, we artificially broaden our momentum resolution by summing spectra over momenta up to 0.3~\AA$^{-1}$, weighting the data to account for the circular selection aperture used in those studies. The resulting spectrum is shown in Fig.~\ref{fig:measurements_overview}(b).

In the energy loss range 0.3-1.5~eV, our results show rough qualitative agreement with these earlier studies. In both cases, no distinct plasmon is observed; rather, a shoulder on the quasielastic line is seen. In our summed spectra, this feature arises from integration over a finite momentum range, and increasing the integration range further suppresses the plasmon feature. Interestingly, both older datasets exhibit a pronounced onset above 1.5~eV, which is absent in all other measurements. The origin of this feature, which seems to be specific to focused-beam experiments, remains unclear.

The agreement in the low-energy range in Figs.~\ref{fig:measurements_overview}(b) arises, in part, because the quasielastic peak was retained in both of these early studies. Notably, Terauchi (1995) \cite{Terauchi1995} demonstrated that subtracting the elastic line from these spectra produced a plasmon-like feature that was not present in the raw data. A key takeaway from these studies is that elastic-line subtraction can give the appearance of a plasmon even when none exists in the measured spectra.

\section{Conclusions}
We report the first transmission EELS study of the charge response of Bi-2212 that combines high energy resolution ($\Delta E \approx$ 39~meV) with high momentum resolution ($\Delta q \approx 0.014$\AA$^{-1}$). To address historical reproducibility issues in this field, we repeated the experiment ten times on five different Bi-2212 flakes (Table~\ref{table:samples}), validated the setup using a standard aluminum reference (Fig.~\ref{fig:metal_strangemetal_comparison}(a)), and made a quantitative comparison with previous studies dating back to 1989. A broad, plasmon-like feature is observed at $q < 0.15$~\AA$^{-1}$; however, we do not observe the RPA-like dispersion reported by Nücker et al. (1989) \cite{Nucker1989}. Instead, the plasmon simply damps out at larger momenta. The pronounced plasmon reported at large $q$ in that early study may be an artifact of subtraction of the zero-loss line, as previously noted by the Tohoku group \cite{Terauchi1995}. 

The absence of a propagating plasmon mode makes the dynamic charge response of Bi-2212 markedly different from that of a Lindhard metal, as illustrated in Fig.~\ref{fig:metal_strangemetal_comparison}. At small momenta ($q < 0.15$~\AA$^{-1}$), a highly damped excitation is observed near 1~eV which is broad, but evident in the raw data. Its intensity decreases with increasing momentum until it vanishes above $q \sim 0.15$~\AA$^{-1}$, where only a featureless continuum remains. This behavior is broadly consistent with reflection M-EELS measurements \cite{Mitrano2018,Husain2019,Chen2024}, though the momentum cutoff differs slightly---0.08~\AA$^{-1}$ in M-EELS versus 0.15~\AA$^{-1}$ in the present study. The energy dependence of the continuum also differs between reflection- and transmission-EELS, indicating that further work is needed to reconcile the two geometries.

The lineshape of the feature observed with T-EELS at small momenta is broadly consistent with resonant inelastic x-ray scattering (RIXS) experiments on Bi-based cuprates \cite{Nag2020,Nakata2025}. RIXS measurements are restricted to $q_z > 0$, where the Fetter model predicts acoustic-like dispersion \cite{Fetter1974}, whereas our T-EELS measurements are performed at $q_z = 0$, where the dispersion is expected to be optical. Consequently, some differences in excitation energies between the two techniques are expected. Although no RIXS experiments have been reported on bilayer Bi-2212, measurements on single-layer Bi-2201 \cite{Nag2020} and trilayer Bi-2223 \cite{Nakata2025} reveal a highly damped plasmon with linewidths exceeding $\sim$0.5~eV, similar to our observations. These results indicate that T-EELS and RIXS are consistent in showing that Bi-based cuprates are incoherent metals with strongly damped plasmons, in contrast to the sharper excitations of a Fermi liquid, as illustrated in Fig.~\ref{fig:metal_strangemetal_comparison}.

One limitation of the present T-EELS experiments is that, unlike M-EELS \cite{Chen2024}, we were unable to access momenta small enough to reach the optical limit ($q < 0.05$~\AA$^{-1}$), where the spectra are obscured by the tails of the bright-field disk. Further work is therefore needed to assess the quantitative consistency of T-EELS measurements with infrared-optics studies, such as those shown in Fig.~\ref{fig:measurements_overview}(c).

\section{Acknowledgments}
We gratefully acknowledge M. Terauchi, J. Zaanen, E. Huang, B. Bradlyn, and P.W. Phillips for helpful discussions, and D. van der Marel for supplying infrared data. 
This work was supported by the Center for Quantum Sensing and Quantum Materials, an Energy Frontier Research Center funded by the U.S. Department of Energy (DOE), Office of Science, Basic Energy Sciences
(BES), under award no. DE-SC0021238. Momentum resolved transmission EELS measurements at the Center for Nanophase Materials Sciences (CNMS) at Oak Ridge National Laboratory were supported by UT-Battelle, LLC under DOE BES contract no. DE-AC05-00OR22725, as well as by the Laboratory Directed Research and Development Program of ORNL.
Some sample preparation instrumentation was supported by NSF MRSEC award DMR-2309037.
P.A. acknowledges additional support from the EPiQS program of the Gordon and Betty Moore
Foundation under Grant No. GBMF9452. 

\bibliography{references}

\end{document}